\documentclass[aps,prd,preprintnumbers,nofootinbib,floatfix,floats,groupedaddress,twocolumn]{revtex4}

\usepackage{bm}
\usepackage[utf8]{inputenc}
\usepackage{latexsym}
\usepackage{dcolumn}
\usepackage{amsmath,amsfonts,amssymb,mathtools}
\usepackage{empheq}
\usepackage{graphicx,epsfig}
\usepackage{float}
\usepackage{color}
\usepackage[active]{srcltx}
\usepackage[caption=false]{subfig}
\usepackage{slashed}
\usepackage{hyperref}
\usepackage{comment}
\usepackage{tikz}
\usetikzlibrary{shapes,snakes}
\usepackage{multirow}
\usepackage[normalem]{ulem}
\usepackage[mathscr]{euscript}
\usepackage{mathrsfs}
\usepackage{enumitem}
\def\nn{\nonumber}
\def\n{\nabla}
\def\pa{{\partial}}
\def\l{\left}
\def\r{\right}
\def\DM{\mathrm{d}}

\newcommand{\gae}{\lower 3pt \hbox{$\,\, \buildrel {\scriptstyle >}\over {\scriptstyle
\sim}\,\,$}}
\newcommand{\lae}{\lower 2pt \hbox{$\, \buildrel {\scriptstyle <}\over {\scriptstyle
\sim}\,$}}

\DeclarePairedDelimiter\bra{\langle}{\rvert}
\DeclarePairedDelimiter\ket{\lvert}{\rangle}
\DeclarePairedDelimiterX\braket[2]{\langle}{\rangle}{#1\,\delimsize\vert\,\mathopen{}#2}

\reversemarginpar

\def\eq#1{{Eq.~(\ref{#1})}}
\def\eqs#1{{Eqs.~(\ref{#1})}}
\def\fig#1{{Fig.~\ref{#1}}}

\begin{document}


\title{Rotating detectors in dS/AdS spacetimes}

 \author{Hari K}
 \email{harik@physics.iitm.ac.in}
 \author{Dawood Kothawala}
 \email{dawood@iitm.ac.in}
 \affiliation{Centre for Strings, Gravitation and Cosmology, Department of Physics, Indian Institute of Technology Madras, Chennai 600 036, India}

\date{\today}
\begin{abstract}
\noindent
We analyse several aspects of detectors with uniform acceleration $a$ and uniform rotation $\Omega$ in de Sitter ($\Lambda>0$) and anti-de Sitter ($\Lambda<0$) spacetimes, focusing particularly on the periodicity, in (Euclidean) proper time $\tau_{\rm traj}$, of geodesic interval $\tau_{\rm geod}$ between two events on the trajectory. For $\Lambda<0$, $\tau_{\rm geod}$ is periodic in ${\rm i} \tau_{\rm traj}$ for specific values of $a$ and $\Omega$. These results are used to obtain numerical plots for the response rate $\dot{\mathcal{F}}$ of Unruh-DeWitt detectors, which display non-trivial combined effects of rotation and curvature through the dimensionless parameter $\Lambda c^2/\Omega^2$. In particular, periodicity does not imply thermality due to additional poles in the Wightman function away from the imaginary axis. We then present some results for stationary rotational motion in arbitrary curved spacetime, as a perturbative expansion in curvature. 
\end{abstract}

\pacs{04.60.-m}
\maketitle
\vskip 0.5 in
\noindent
\maketitle
\section{Introduction} \label{sec:intro} 

Studying physical processes in a uniformly accelerated frame of reference in Minkowski spacetime\cite{Fulling:1972md,Davies:1974th,Unruh:1976db} often forms a first step towards understanding these processes in curved spacetime. The mapping between these is provided by the so called \textit{principle of equivalence}. However, by its very nature, the equivalence principle gives no insights into the quantitative role of curvature in these processes. One can obtain curvature corrections perturbatively, of course, but such results are of limited interest from a conceptual point of view since the limit of zero acceleration can not then be taken. A well known result\cite{deser1997accelerated,deser1998equivalence,deser1999mapping} that illustrates this is the response of a uniformly accelerated Unruh-DeWitt detector in de Sitter spacetime with cosmological constant $\Lambda$, which is thermal with a temperature proportional to $\sqrt{a^2 + \Lambda}$. A perturbative analysis would yield the leading curvature correction at $O(\Lambda/a^2)$, and miss the $a \to 0$ limit. 

In a recent work \cite{hari:2021gns}, it was shown that similar contribution due to (electric part of) Riemann tensor arises in general curved spacetimes, of which the (anti-) de Sitter results are but special cases. A remarkable partial summation of an infinite series once again yields new insights that one could not have gained from a naive application of the equivalence principle.

In this paper, we broaden the scope of such results by considering uniformly accelerating and rotating detectors in maximally symmetric spacetimes and obtaining certain exact results that should be relevant for the response of such detectors. We also give the result for arbitrary curved spacetimes, but, unfortunately, we are forced to leave it as a perturbation expansion since we have not been able to re-sum even a subset of terms in presence of rotation. Nevertheless, the exact results we present for rotating detectors in dS/AdS spacetimes by themselves yield important insights into the non-trivial of $\Lambda$ (including its sign) on the response of the detectors.

Three key results derived in this work are:

\begin{enumerate}[wide, labelwidth=!, labelindent=0pt, itemsep=0.5cm]
\item \textbf{Relation between proper time and geodesic interval:} Remarkably, the relation between the geodesic distance, $\Delta\tau_{\rm geod}$ and proper time distance, $\Delta\tau_{\rm traj}$ for a stationary trajectory with constant acceleration and constant torsion in a maximally symmetric spacetime can be expressed in a form closely resembling the structure in Minkowski spacetime (see \eqs{sigma2-ms-2}

\item \textbf{Periodicity in $\Delta \tau_{\rm geod}^2$:} The above result yields the condition for periodicity of $\Delta \tau_{\rm geod}^2$ in Euclidean proper time ($\Delta\tau_{\rm traj}\rightarrow i \Delta t_{\rm traj}$) for $\Lambda<0$, for specific values of acceleration and torsion.

\item \textbf{General spacetime:} We also obtain a perturbative expression for $\Delta \tau_{\rm geod}^2$ for stationary trajectories in arbitrary curved spacetime, assuming derivatives of Riemann to be small.

\end{enumerate} 

The (square of) geodesic interval, $\sigma(x,x^{\prime})^2=-\Delta \tau_{\text{geod}}^2(x,x\prime)$ between two points on a trajectory is an important quantity that appears prominently in the analysis of classical as well quantum measurement processes in curved spacetimes. The latter, for instance, through its appearance in the leading short distance (Hadamard) form of the two-point function
        \begin{eqnarray}
        G_{\rm H}(x,y) = \frac{{\Delta^{1/2}(x,y)}}{\sigma(x,y)^2} + \ldots
        \end{eqnarray}
where $\Delta(x,y)$ is the van Vleck determinant, derived again from $\sigma^2$. Amongst many things, the above form serves as the key mathematical tool to evaluate the response of an Unruh-DeWitt detector coupled to quantum field in an arbitrary curved spacetime.

\textit{NB:} The curvature constant (cosmological constant) used for de Sitter in this paper, $\Lambda$, is conventionally denoted by $\Lambda/3$ in the context of cosmology.

\section{Stationary motion}
The trajectory of a timelike curve is specified using functions $x^a(\tau)$ defined at each point of the curve, with $\tau$ as the proper time along the curve. A more elegant way to represent the trajectory is through its curvature invariants, which, in four dimensions, are acceleration, torsion, and hypertorsion. At every point on the trajectory an orthonormal tetrad, $e^{\textsf{i}}_{a}$ can be constructed from the derivatives of $x^{a}(\tau)$, which satisfy the condition,
\begin{eqnarray}
e^{\textsf{i}}_{a}e_{\textsf{i}b}=\eta_{ab}
\end{eqnarray}
Here, $\eta_{ab}=\text{diag}(-1,1,1,1)$. These tetrads serve as the basis vectors for the vector space at each point on the worldline. They obey the Serret-Frenet equations given by,
\begin{eqnarray}
\frac{D e^{\textsf i}_{a}}{\DM \tau}=K_{a}^{\phantom{a}b} e^{\textsf{i}}_{b}
\end{eqnarray}
where the structure of the matrix $K_{ab}$ is given by,
\begin{eqnarray}
    K_{ab}=\begin{bmatrix}
    0 & -a(\tau) & 0 & 0 \\
    a(\tau) & 0 & \Omega(\tau) & 0 \\
    0 & -\Omega(\tau) & 0 & \lambda(\tau) \\
    0 & 0 & -\lambda(\tau) & 0 \\
    \end{bmatrix}
\end{eqnarray}
where $a(\tau)$ is the magnitude of acceleration, $\Omega(\tau)$ is the torsion, and $\lambda(\tau)$ is the hypertorsion. These are the curvature invariants of the trajectory. Note that the matrix $K_{ab}$ is antisymmetric, $K_{ab}=-K_{ba}$. The worldline is stationary when these invariants are constant and do not depend on the parameter $\tau$. These stationary worldlines are classified into six categories (see Ref. \cite{letaw1981stationary,letaw1982stationary,letaw1981quantized}) for Minkowski spacetime according to the values of curvature invariants. 

For a stationary trajectory, let $u^i(\tau)$ be the 4-velocity and $n^i(\tau)$ be the unit vector along the direction of 4-acceleration. We will be using Serret-Frenet equations to construct the tetrads and will express the worldline in terms of the tetrads at $\tau=0$. One starts with the tangent vector to the curve, and a unit vector in the direction of acceleration, from which remaining orthogonal vectors are obtained using Gram-Schmidt orthogonalization.
\begin{eqnarray}
\nabla_{\bf u} u^k &=& a\, n^k \; ; \quad \nabla_{\bf u} n^k = \Omega\, b^k+a\, u^k\; ;
\nn \\
\quad \nabla_{\bf u} b^k &=& \lambda\, d^k-\Omega\, n^k\; ; \quad \nabla_{\bf u} d^k =-\lambda\, b^k
\label{serret-frenet}
\end{eqnarray}
Here, $\nabla_{\bf u}$ is the covariant derivative,  $n^k=a^k/a$ is the normal vector in the direction of acceleration, $b^k$ is the binormal orthogonal to $u^k$ and $a^k$, and $d^k$ is another unit vector orthogonal to all other unit vectors. These equations are more succinctly expressed in terms of the Fermi derivative defined by,
\begin{eqnarray}
\frac{D_F Y^i}{\DM \tau}=\nabla_{\bf u} Y^i+ \Omega^i_{\phantom{i}k} Y^k=0\nn
\end{eqnarray}
where, $\Omega^{ik}:=a^i u^k-u^i a^k+\omega^{ik}$ and $\omega^{ik}:=\varepsilon^{ikjl} u_j \omega_l $. For stationary motion, $K_{ab}=\Omega_{ba}$.

The key geometrical quantity, which will also be our main focus, is the geodesic distance between two points on a stationary trajectory characterized by Serret-Frenet equations. To do this, we will follow the method sketched in \cite{hari:2021gns}. This method essentially uses 
Riemann normal coordinates (RNC) to solve for the trajectory as a power series
\begin{eqnarray}
x^i(\tau)=\sum_{n=0}^{\infty}\frac{\tau^n}{n!}\l[\frac{\DM^n x^i}{\DM \tau^n}\r]_{\tau=0}
\label{traj-gen-st}
\end{eqnarray}
in which the coefficients on the RHS are determined by taking higher derivatives of the defining equation 
\begin{eqnarray}
\frac{\DM^2 x^i}{\DM \tau^2}+\Gamma^{i}_{bc}\frac{\DM x^b}{\DM \tau}\frac{\DM x^c}{\DM \tau}=a^i
\label{acc-eq}
\end{eqnarray}
and using the Serret-Frenet equations. This method was explained and used in \cite{hari:2021gns} for rectilinear, uniformly accelerated motion in arbitrary curved spacetime, and was shown to yield a remarkable result involving an analytically resummable piece of an otherwise infinite (perturbative) expansion. In this work, we will explore the effects of rotation to see if a similar result holds in this case.

\section{Maximally symmetric spacetimes}

As a first step towards studying stationary trajectories in arbitrary curved spacetime, we will consider stationary trajectories in maximally symmetric spacetimes. As we will show, some very interesting analytic results can be obtained in this case, with several remarkable similarities and mappings with the corresponding results in Minkowski spacetime, which we have discussed in Appendix \ref{stat-mink}; we will refer to the results in this appendix after deriving the corresponding results in maximally symmetry.

We will mostly use the metric of maximally symmetric spacetimes in embedding coordinates $X^i$, given by\cite{weinberg1972gravitation},
\begin{eqnarray}
g_{ab}=\eta_{ab}+\frac{\Lambda}{1-\Lambda\eta_{ij}X^iX^j}\eta_{ac}\eta_{bd}X^cX^d
\end{eqnarray}
Then the Christoffel connections can be expressed as $\Gamma^{a}_{bc}=\Lambda X^ag_{bc}$. It can be easily shown that the stationary motion with acceleration and torsion will be always in the hyperplane $u^i(0)$-$n^i(0)$-$b^{i}(0)$ using Serret-Frenet equations. 

\subsection{Trajectory and geodesic distance}
Further, using \eq{serret-frenet} and \eq{acc-eq} the Taylor series expansion for the trajectory similar to Minkowski spacetime can be obtained as,
\begin{eqnarray}
\frac{\DM X^i}{\DM \tau}&=&U^i\nonumber\\
\frac{\DM^2 X^i}{\DM \tau^2}&=&a N^i +\Lambda X^i\nonumber\\
\frac{\DM^3 X^i}{\DM \tau^3}&=&a\, \Omega\, B^i + (a^2+\Lambda) \frac{\DM X^i}{\DM \tau}\nonumber\\
\frac{\DM^{\mathsf{p}} X^i}{\DM \tau^{\mathsf{p}}}&=&(a^2-\Omega^2+\Lambda) \frac{\DM^{\mathsf{p}-2} X^i}{\DM \tau^{\mathsf{p}-2}}+\Omega^2 \Lambda  \frac{\DM^{\mathsf{p}-4} X^i}{\DM \tau^{\mathsf{p}-2}}\nonumber
\end{eqnarray}
Here, $\mathsf{p}\ge 4$, $U^i$, $N^i$, $B^i$ are similar to unit vectors $u^i$, $n^i$, $b^i$ in Minkowski spacetime, but now defined using the embedding coordinates. The above equations are obtained by expanding the covariant derivatives and using the Christoffel connections previously described. Solving the recursion equation will determine all the higher-order terms. The terms in each direction can be summed into hyperbolic functions, $\sinh$ and $\cosh$. The final form of the summed series expressing the trajectory is,
\begin{widetext}
\begin{eqnarray}
Z^i(\tau)&=&\frac{a}{q_0^2} \left[\cosh\left(q_{+}\tau\right)-\cosh\left(q_{-}\tau\right)\right] \, N^i(0) - 
\frac{ a\,\Omega}{q_+ q_- q_0^2}\left[
q_{+} \sinh\left(q_- \tau\right) - 
q_{-} \sinh\left(q_+\tau\right)\right]\, B^i(0) 
\nonumber \\
 && - \frac{1}{q_+q_-q_0^2}\left[(q_-^2+\Omega^2)q_+\sinh\left(q_-\tau\right)-(q_+^2+\Omega^2)q_-\sinh\left(q_+\tau\right)\right]\, U^i(0) \nn \\
&& +\frac{\Lambda}{q_0^2}\l[\l(\frac{q_{+}^{2}+\Omega^2}{q_+^2}\r)\cosh(q_+\tau)-\l(\frac{q_{-}^{2}+\Omega^2}{q_-^2}\r)\cosh(q_-\tau)\r]X^i_0
\label{traj-ms}
\end{eqnarray}
\end{widetext}
Here, $q_0:=\left[(a^2-\Omega^2+\Lambda)^2+4\Lambda \Omega^2\right]^{1/4}$, $q_{\pm}:=(1/\sqrt{2})\sqrt{(a^2-\Omega^2+\Lambda) \pm q_0^2}$ and $X_0^{i}$ is the initial position. These constants are similar to the ones obtained for stationary motion with all curvature invariants. In fact, there seems to be mapping between these trajectories, which will be discussed soon.

The geodesic interval can be found using the relation between the arc-length and chordal distance in maximal symmetric spacetimes. The chordal distance is evaluated using $\eta_{ab}\Delta Z^a(\tau)\Delta Z^b(\tau)$, where $\Delta Z^a(\tau)=Z^{a}(\tau)-Z^a(0)$ and arrived at the relation for geodesic interval as,
\begin{eqnarray}
\Delta\tau_{\rm geod}^2=-\frac{1}{\Lambda}\left[\sin^{-1}\left(\sqrt{-\Lambda \Delta\tau_{\rm ms}^2}\right)\right]^2
\label{sigma2-ms-1}
\end{eqnarray}
Here, $\Delta\tau_{\rm ms}^2:=-(\eta_{ab}\Delta Z^a \Delta Z^b)$ given by,
\begin{widetext}
\begin{eqnarray}
    \Delta\tau_{\rm ms}^2&=&\Delta\widetilde{\tau}^2\left(1+\frac{1}{4}\Lambda\Delta\widetilde{\tau}^2\right) 
    \label{sigma2-ms-3}\\
    \Delta\widetilde{\tau}^2 &=& \frac{4}{q_0^2}\left[\left(1+\frac{\Omega^2}{q_+^2}\right)\sinh^2\left(\frac{q_+ \Delta\tau_{\rm traj}}{2}\right) -\left(1+\frac{\Omega^2}{q_-^2}\right)\sinh^2\left(\frac{q_- \Delta\tau_{\rm traj}}{2}\right)\right] 
    \label{sigma2-ms-4}
\end{eqnarray}    
\end{widetext}
Equation \ref{sigma2-ms-1} can be rearranged using \eq{sigma2-ms-3} to a similar form as obtained in Eq. (23) of \cite{hari:2021gns},
\begin{align}
    \Delta\tau_{\rm geod}^2=\frac{4}{\Lambda}\left[\sinh^{-1}\left(\frac{1}{2}\sqrt{\Lambda \Delta\widetilde{\tau}^2}\right)\right]^2
\label{sigma2-ms-2}
\end{align}
The equation for the hyperbolic trajectory in the ref. \cite{hari:2021gns} will be a special case when $\Omega=0$. Note that the same relation can also be arrived at by using the mapping between the embedding coordinates and the RNC\cite{hari2020normal}, $X^i=\Delta^{-1/(D-1)}\hat{x}^i$, where, $\Delta=\Delta(0,\hat x)$ is the van Vleck determinant with $D$ as the dimension of spacetime. The expression, $-\Delta\tau_{\rm geod}^2=\eta_{ab}\hat{x}^a\hat{x}^b$ can be used to evaluate the geodesic interval.

\subsection{Periodicity in $\Delta \tau_{\rm geod}^2$}\label{sec:periodicity}

The periodicity in $\Delta \tau_{\rm geod}^2$ for Euclidean time($\Delta\tau_{\rm traj}\rightarrow i \Delta t_{\rm traj}$) can be analysed using \eq{sigma2-ms-4}. Equation \ref{sigma2-ms-4} will be periodic only when both conditions given below are satisfied.

i) $\{q_+,q_-\}\in \mathbb{R}$

ii) $q_+/q_- \in \mathbb{Q}$

The first condition restricts $q_-\in \mathbb R$, which implies
\begin{eqnarray}
1-\sqrt{1+\frac{4\Lambda \Omega^2}{a^2-\Omega^2+\Lambda}}>0,\quad \sqrt{a^2-\Omega^2+\Lambda}>0\nn
\end{eqnarray}
The above inequalities are only satisfied for $\Lambda < 0$ and $a^2>\Omega^2+|\Lambda|$. 

Using the first condition, the second condition can be simplified as,
\begin{eqnarray}
    \frac{q_+}{q_-}=\frac{\sqrt{1+\sqrt{1-\frac{4|\Lambda|\Omega^2}{(a^2-\Omega^2-|\Lambda|)^2}}}}{\sqrt{1-\sqrt{1-\frac{4|\Lambda|\Omega^2}{(a^2-\Omega^2-|\Lambda|)^2}}}}=r \nn \\
    \implies \frac{|\Lambda|\Omega^2}{(a^2-\Omega^2-|\Lambda|)^2} = \frac{ r^2}{(r^2+1)^2}
\end{eqnarray}
Here, $r\in \mathbb{Q}$ and $r>1$. The above condition ensures that $q_+$ and $q_-$ are commensurable. Therefore, the conditions for periodicity in $\Delta\tau_{\rm geod}^2$ can be summarized as:

\begin{eqnarray}
    \Lambda &<& 0 \nn \\
    a^2 &>& \Omega^2 + |\Lambda| \nn \\
    \frac{|\Lambda|\Omega^2}{(a^2-\Omega^2-|\Lambda|)^2} &=& \frac{ r^2}{(r^2+1)^2} \nn
\\ \nn
\end{eqnarray}

\subsection{Mappings between stationary motions in Minkowski and dS/AdS.}

1. \underline{\emph{Helical motion in Minkowski and rotation in dS/AdS}}:

The RHS of \eq{sigma2-ms-4} and \eq{mink-sigma-sqr} have a striking similarity. The functional form of these equations are similar. If we try compare the constants, the mapping between these equations can be summarized as,

\begin{center}
\begin{tabular}{c  c} 
\hline
Minkowski & \quad Maximally symmetric \\ [0.8ex] 
\hline\hline
$a_{\eta}^2$ & $a^2+\Lambda$  \\ [0.8ex] 
$\Omega_{\eta}^2$ & $a^2(\Omega^2/(a^2+\Lambda))$  \\ [0.8ex] 
$\lambda^2$ & $\Lambda(\Omega^2/(a^2+\Lambda))$  \\ [0.8ex] 
\hline
\end{tabular}
\end{center}

Note that $\Omega_{\eta}^2+\lambda^2=\Omega^2$. The curvature constant of maximally symmetric spacetime, $\Lambda$, plays a similar role of hypertorsion, $\lambda$ in flat spacetime.

2. \underline{\emph{Rotation in Minkowski and rotation in dS/AdS with $a= \Omega$}}:

In the case of maximally symmetric spacetimes, when the acceleration and torsion are equal, \eq{sigma2-ms-3} reduces to,
\begin{equation}
    \Delta\tau_{\rm ms}^2=\frac{4(\Lambda+\Omega^2)}{\Lambda^2}\sinh^2\left(\frac{\sqrt{\Lambda}}{2}\Delta\tau_{\rm traj}\right)-\frac{\Omega^2\Delta\tau_{\rm traj}^2}{\Lambda}
\end{equation}
The mapping between the flat spacetime and maximally symmetric spacetime for this motion is, $a_{\eta}^2\rightarrow a^2+\Lambda$ and $\Omega_{\eta} \rightarrow a=\Omega$.

\section{Route to generalise the results to arbitrary curved spacetimes}
We now turn focus on arbitrary curved spacetimes, hoping again to obtain the relation between the geodesic interval and the proper time interval for points on stationary trajectories in these spacetimes, characterised again through Serret-Ferret conditions. The method we will employ is the one based on a judicious use of RNC, described in Sec. II of Ref. \cite{hari:2021gns}. The RNC is setup at some initial point $p_0$ and the trajectory of the probe at the point $p$ is expressed using \eq{traj-gen-st}. From the definitions of RNC we have, $\hat{z}^a(\Delta \tau)=\hat{x}^{a}(p)=\l(\Delta \tau_{\text{geod}}\r)\hat{t}^a(0;\Delta\tau)$ and $(\Delta \tau_{\text{geod}})^2 = \eta_{ab} {\widehat z}^a(\Delta \tau) {\widehat z}^b(\Delta \tau)$, where $\hat{t}^a$ is the tangent vector along the geodesic curve connecting the points $p_0$ and $p$. Utilizing \eq{acc-eq} and its higher derivatives in RNC along with Serret-Frenet equations, each terms in the series can be obtained as,
\begin{eqnarray}
    \dot{\widehat{z}^i}(0)&=&u^i \rvert_{p_0} = u^i(0) \nn \\ 
    \ddot{\widehat{z}^i}(0)&=&\left[ a\,n^i-\Gamma^i_{m k}u^m u^k\right]_{p_0}=\left[a \, n^i \right]_{p_0} \nn \\
    \dddot{\widehat{z}^i}(0) &=& \left[a\,u^k \pa_k n^i - \Gamma^i_{m j, k}u^k u^m u^j -2 \Gamma^i_{m j}(u^k\pa_ku^j)\right]_{p_0} \nn \\
    &=& \left[a\,\n_{\bm u} n^i - 3 \Gamma^i_{k m}u^ka^m - \Gamma^i_{m j,k}u^k u^m u^j \r. \nn \\
    && \l. + \Gamma^i_{m j}\Gamma^{j}_{k l}u^m u^k u^l \right]_{p_0} \nn \\
    &=& \left[a\,\n_{\bm u} n^i - \Gamma^i_{m j,k}u^k u^m u^j \right]_{p_0} \nn \\
    &=& \l[a\,\Omega\, b^i + a^2\, u^i - \Gamma^i_{m j,k}u^k u^m u^j \r]_{p_0} \nn
\end{eqnarray}
The Christoffel connections in the above terms are evaluated using the series expansions in RNC. In the case of non-rotating motion ($\Omega=0$), this procedure gave the following series in Ref. \cite{hari:2021gns},
\begin{align}
    \Delta\tau_{\rm geod}^2 &=\tau^{2}_{\rm acc} + \frac{1}{12}a^{2} {\tau}^{4}_{\rm acc} + \frac{1}{360}\left( a^{4} + 3 a^2 {\mathscr E}_n \right) {\tau}^{6}_{\rm acc}\nn
    \\
    & \quad + \frac{1}{20160}\left(a^{6} + 17 a^2 {\mathscr E}_n^2 + 18 a^{4} {\mathscr E}_n \right) {\tau}^{8}_{\rm acc}  \nn
    \\
    & \quad + \frac{1}{1814400} \left( a^{8} + 81 a^{6} {\mathscr E}_n +  339 a^{4} {\mathscr E}_n^2 \right. \nn \\
    & \left. \quad + 155 a^2 {\mathscr E}_n^3 \right) {\tau}^{10}_{\rm acc} + O(\tau^{12}_{\rm acc})+ \mathscr{R}_A + \nabla R\text{ terms}
    \label{eq:sigma-tau-relation}
\end{align}
Here, ${\mathscr E}_n := R_{0n0n}=R_{abcd} u^a n^b u^c n^d$ and $\mathscr{R}_A$ collectively represents all terms that have at least one Riemann tensor with at least one index which is neither $0$ nor $n$. A remarkable feature of this series, as was pointed out in \cite{hari:2021gns}, is that it admits a partial re-summation involving terms containing acceleration and the tidal part of the Riemann tensor into a nice analytic function, 
\begin{eqnarray}
    \Delta\tau_{\rm geod} &=& \frac{2}{\sqrt{{-\mathscr E}_n}} \sinh ^{-1}\Biggl[\sqrt{\frac{-{\mathscr E}_n}{a^2-{\mathscr E}_n}} \sinh \left(\frac{\sqrt{a^2-{\mathscr E}_n} \, \tau_{\rm acc} }{2} \right)     \Biggl] \nn \\
    & & + \mathscr{R}_A  + \nabla R\text{ terms}
    \label{eq:sigma2-general}
\end{eqnarray}
A discussion on the off-plane components of Riemann tensor terms ($\mathscr{R}_A$) and derivatives of Riemann tensor terms ($\nabla R$ terms) are included in Appendix \ref{sec:exp-general}. 

For rotational motion $\Omega \neq 0$, we can obtain a similar series expansion in curvature using \texttt{CADABRA}. To $O(\Delta\tau_{\text{traj}}^9)$, this is given by
\begin{widetext}
\begin{eqnarray}
    \Delta \tau_{\rm geod}^2 &=& \Delta \tau_{\rm traj}^2 + \frac{1}{12} a^2 \Delta \tau_{\rm traj}^4 + \frac{1}{360} a^2(a^2 -\Omega^2 +3 R_{0 n 0 n}) \Delta \tau_{\rm traj}^6 - \frac{1}{120}
    a^2 \Omega\, R_{n 0 b 0}\, \Delta \tau_{\rm traj}^7 + \frac{1}{20160} \Bigg( a^6 - 2 a^4 \Omega^2 
    \nn \\
    &&  + a^2\Omega^4 + 18 a^4 R_{0 n 0 n} + \frac{128}{3} a^2 \Omega^2 R_{0 b 0 b} - 17a^2 R_{0 n 0 \bullet}R^{\bullet}_{0 n 0} + 50 a^2 \Omega^2 R_{0n0n} + 32 a^3 \Omega R_{n b n 0} \Bigg) \Delta \tau_{\rm traj}^8 + O(\Delta \tau_{\rm traj}^9) \nn \\
\end{eqnarray}
\end{widetext}
Here, $\bullet$ index represent the direction along acceleration and torsion. However, unlike the case with $\Omega=0$, we have not been able to obtain even a partial re-summation of this series to a nice analytic function of acceleration, torsion, and different Riemann tensor components. The exact result in dS/AdS, which was helpful in partially identifying such a re-summation in Ref. \cite{hari:2021gns}, is not of much use for $\Omega \neq 0$ since the presence of an extra spacelike direction introduces additional components of Riemann, which are all equal or zero in maximal symmetry. 

\section{Comment on rotating observers in BTZ black hole}

In Ref. \cite{Hodgkinson:2012mr}, the authors discuss about the detector response of an observer who is corotating with the horizon of Bañados-Teitelboim-Zanelli (BTZ) blackhole\cite{Banados:1992wn,Banados:1992gq,carlip19952+}. The BTZ blackhole metric can be constructed by a suitable coordinate transformation from 2+1 Anti-de Sitter($\text{AdS}_3$) spacetime. The $\text{AdS}_3$ metric is given by,
\begin{eqnarray}
    d\,s^2=-\l(\frac{\hat{r}^2}{\ell^2} -1\r)d\,\hat{t}^2+\l(\frac{\hat{r}^2}{\ell^2} -1\r)^{-1}d\,\hat{r}^2+r^2 d\,\hat{\phi}^2 \nn \\
\end{eqnarray}
with $f=\hat{r}^2/\ell^2 -1$ and $\ell$ is the curvature length scale. 
By the following identification, 
\begin{eqnarray}
    \hat{t}&=&\frac{1}{\ell}(r_{+}t-r_{-}\ell\phi) \nn \\
    \hat{\phi}&=&\frac{1}{\ell}\l(r_{+}\phi-\frac{r_{-}t}{\ell}\r) \nn \\
    \hat{r}&=&\ell\sqrt{\frac{r^2-r_{-}^2}{r_{+}^2-r_{-}^2}}\nn
\end{eqnarray}
the metric of exterior region of black hole is obtained as,
 \begin{eqnarray}
    d\,s^2=-(N^{\perp})^2d\,t^2+(f)^{-2}d\,r^2+r^2\l(d\phi + N^{\phi} d\,t\r)^2 \nn \\
\end{eqnarray}
where,
\begin{eqnarray}
    N^{\perp}=\l(-M+\frac{r^2}{\ell^2}+\frac{J^2}{4r^2}\r)^{1/2}; \quad N^{\phi}=-\frac{J}{2r^2}
\end{eqnarray}
\\
with $M=(r_{+}^2+r_{-}^2)/\ell^2$ and $J=2r_{+}r_{-}/\ell$. The corotating observer in ref. \cite{Hodgkinson:2012mr}, rigidly rotates with the horizon having an angular velocity, $\omega_{H}=r_{-}/(r_{+}\ell)$. The temperature obtained from the response of the detector for this observer is $T=(1/2\pi)(\alpha-1)^{-1/2}$, where $\alpha=(r^2-r_{-}^2)/(r_{+}^2-r_{-}^2)$. The Seret-Frenet equation for this observer shows zero torsion and the acceleration is constant for $r=\text{constant}$ with a magnitude, $a=(1/\ell)\sqrt{(r^2-r_{-}^2)/(r^2-r_{+}^2)}$. If we use the inverse coordinate transformation from BTZ to AdS$_3$, the temperature will read as, $T=(1/2\pi)(\hat{r}^2-\ell^2)^{-1/2}$ which is exactly equal to the expression for uniformly accelerating observer in anti-de Sitter spacetime\cite{deser1997accelerated,deser1998equivalence,deser1999mapping,hari:2021gns}, $T=(1/2\pi)\sqrt{a^2+\Lambda}$ where $a$ is the magnitude of the acceleration and $|\Lambda|^{-1/2}$ is the curvature length scale. This temperature can be read off from the coefficient of $\tau_{\text{traj}}$ in the geodesic distance if it is invariant for $\tau_{\text{traj}}\rightarrow \tau_{\text{traj}}+2\pi i$.

When one considers an observer rotating with $\omega\ne \omega_{H}$ at some constant radius, $r=r_0$, the observer is in stationary motion with constant acceleration and torsion. The trajectory in BTZ coordinates will be, $(A\,\tau_{\text{traj}},r_0,A\,\omega\,\tau_{\text{traj}})$, where, $A=\l[r^2(1/\ell^2-\omega^2)+J\, \omega -M\r]^{-1/2}$. Using the formula for the geodesic distance in the flat embedding spacetime, $\mathbb{R}^{2,2}$ given in ref. \cite{Hodgkinson:2012mr}, the geodesic distance for the trajectory in flat embedding spacetime for the motion described here can be obtained as,
\begin{widetext}
\begin{eqnarray}
    -\Delta \widetilde{\tau}^2 = 2\,\alpha(r_0)\,\sinh^2\l[\frac{A\, r_{+}}{2\ell} \l(\omega -\omega_{H}\r) \Delta \tau_{\text{traj}} -\frac{r_{+}}{\ell}\pi\,n\r] + 2\l(1-\alpha(r_0)\r)\sinh^2\l[\frac{A\, r_{+}}{2} \l(\frac{1}{\ell^2} -\omega\,\omega_{H}\r) \Delta \tau_{\text{traj}} -\frac{r_{-}}{\ell}\pi\,n\r]
    \label{sigma-btz-rotation}
\end{eqnarray}
\end{widetext}
where, $n\in \mathbb{Z}$. We tried to map this trajectory from BTZ spacetime to $\text{AdS}_3$ spacetime using the inverse coordinate transformation, which gives the trajectory as, 
\begin{eqnarray}
    \hat{t}&=&A\,r_{+}\l(\frac{1}{\ell^2}-\omega\,\omega_H\r)\tau_{\text{traj}} \nn\\
    \hat{r}&=&\sqrt{\alpha}\,\ell \nn \\
    \hat{\phi}&=&\frac{A\,r_{+}}{\ell}\l(\omega-\omega_H\r)\tau_{\text{traj}}\nn
\end{eqnarray}
Estimating acceleration and torsion for this trajectory and plugging into \eq{sigma2-ms-4} gives the same expression as obtained in \eq{sigma-btz-rotation} with $n=0$. For the periodicity in geodesic distance when $\tau_{\text{traj}}\rightarrow i\,t_{\text{traj}}$ the coefficients of $\Delta\tau_{\text{traj}}$ should be commensurate,
\begin{eqnarray}
    \frac{\l(\omega -\omega_{H}\r) }{\ell\l(\frac{1}{\ell^2} -\omega\,\omega_{H}\r)}=m
\end{eqnarray}
where, $m\in \mathbb{Q}$.

\begin{figure*}
\includegraphics[width=5.50cm]{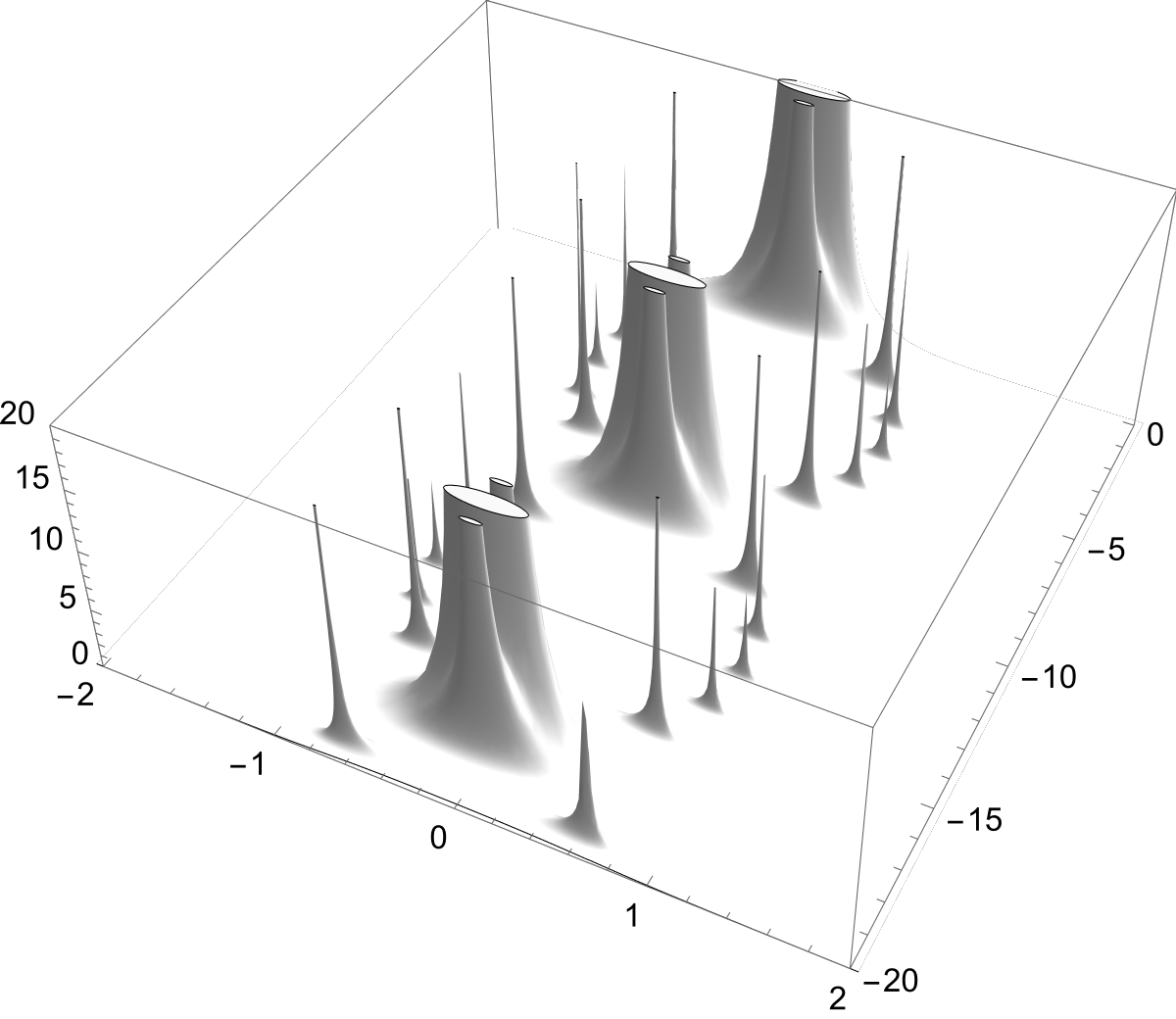}
\hskip 10pt
\includegraphics[width=5.50cm]{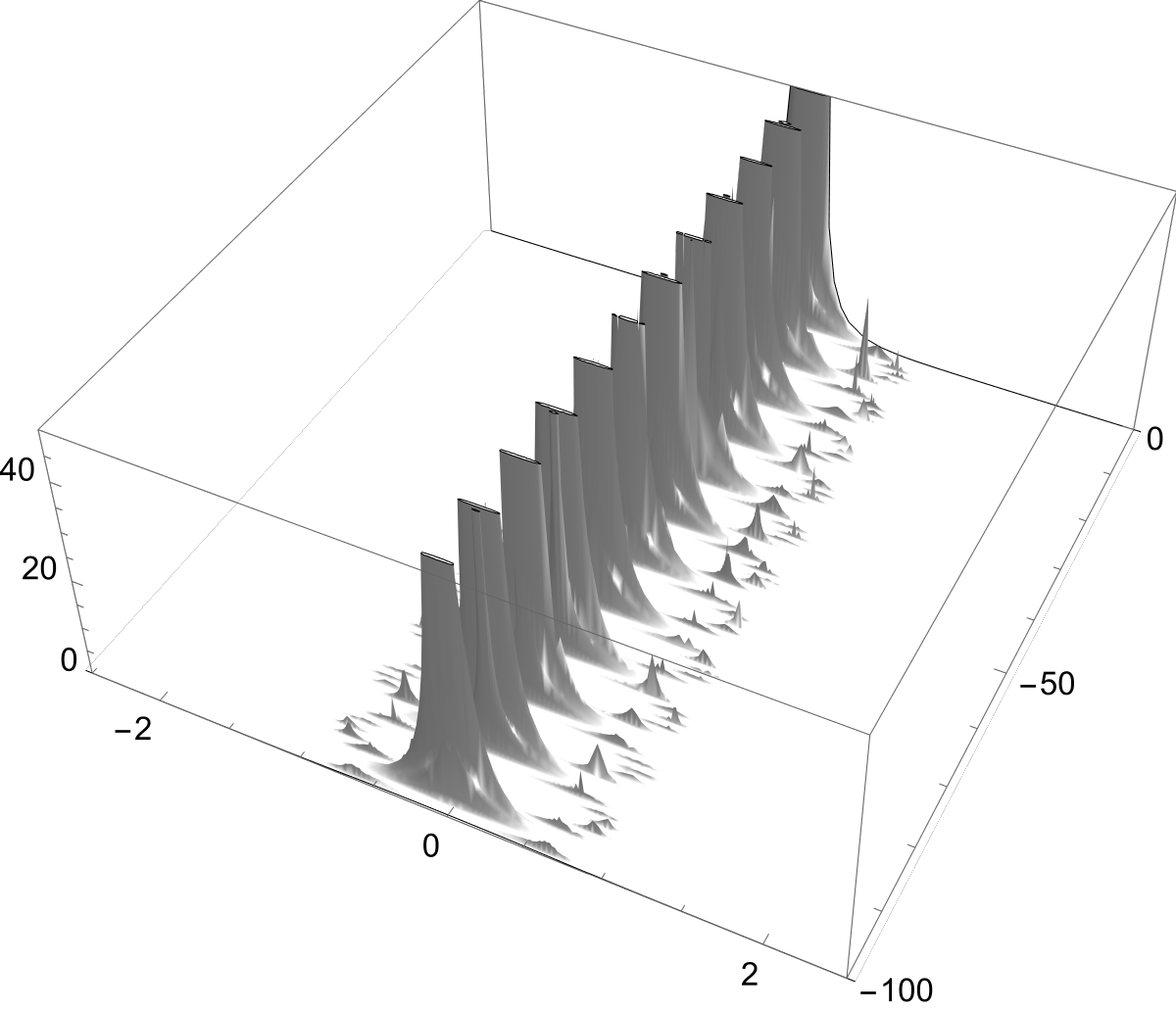}
\hskip 10pt
\includegraphics[width=5.50cm]{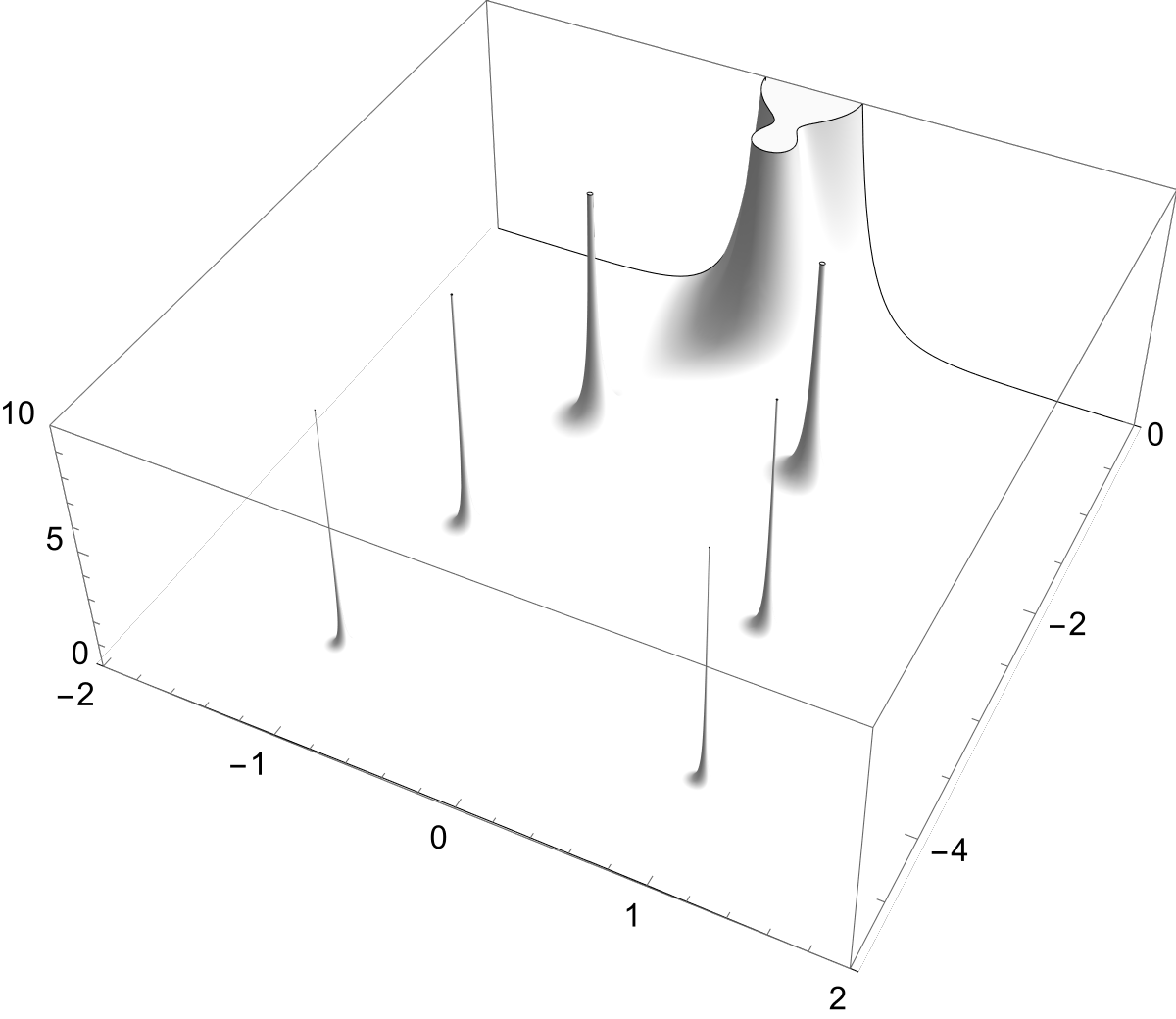}
\caption{The pole structure of $1/\Delta\tau_{\text{geod}}^2$ for $\tau\rightarrow i t$ in i) de Sitter ii) Anti-de Sitter and iii) Minkowski spacetime with $a/\sqrt{|\Lambda|}=1.5\sqrt{17}$ and $\Omega/\sqrt{|\Lambda|}=3.5$. For Anti-de Sitter, the parameters chosen satisfy the periodicity conditions (see text) to demonstrate periodicity of poles on the imaginary axis.}
    \label{fig:pole-structure}
\end{figure*}

\section{Detector response}

Consider a particle detector model\cite{Unruh:1976db,DeWitt:1980hx,book:Birrell} with two energy levels, $\ket{E_0}_d$ and $\ket{E_1}_d$ that moves on a stationary trajectory, $x^i(\tau)$ as discussed in the previous sections. It interacts with the free real scalar field $\phi$ through an interaction Hamiltonian,
\begin{eqnarray}
    H_{\text{int}}=\lambda \, \chi(\tau)\, m(\tau)\, \phi(x^i(\tau))
\end{eqnarray}
where, $\lambda$ is the coupling constant, $\chi(\tau)$ is the switching function, and $m(\tau)$ is the monopole moment. We consider adiabatic switching for the detector. The detector is in its ground state when the interaction with the scalar field begins and the field is in a state, $\Phi$ and assumes that it satisfies the Hadamard property. To the first order in perturbation theory, the probability of the detector to be in the excited state can be expressed as,
\begin{eqnarray}
    P\l(\Delta E\r)=\lambda^2\, \lvert \prescript{}{d}{\bra{E_0}}m(0)\ket{E_1}_{d} \rvert^2 \, \mathcal{F}(\Delta E)
\end{eqnarray}
where $\mathcal{F}(\Delta E)$ is the response function that contains the information on the trajectory of the detector and the initial state of the field, and $\Delta E$ is the energy gap of the detector, $E_1-E_0$. The response function for adiabatic switching is given by,
\begin{eqnarray}
\mathcal{F}\l(\Delta E\r)=\lim_{\epsilon\to0^{+}}\int_{-\infty}^{\infty} \, \DM \tau^{\prime}\, \int_{-\infty}^{\infty}\, \DM \tau\, e^{-i\Delta E(\tau-\tau^{\prime})} \, G^{+}_{\epsilon}\l(\tau,\tau^{\prime}\r) \nn\\
\end{eqnarray}
Here $G^{+}_{\epsilon}\l(\tau,\tau^{\prime}\r)$ is the pull-back of the Wightman function, $G^{+}_{\epsilon}\l(x^{i}(\tau),x^{i}(\tau^{\prime})\r)$ with $i\varepsilon$ prescription. A more useful quantity is the proper time derivative of the response function known as the instantaneous transition rate. For stationary motion, $G^{+}(\tau,\tau^{\prime})\rightarrow G^{+}(\tau-\tau^{\prime})$ and using the transformation $\tau-\tau^{\prime}=u$ and $\tau^{\prime}=v$, the transition rate can be defined as,
\begin{eqnarray}
    \dot{\mathcal{F}}(\Delta E)=\lim_{T\to\infty} \frac{\mathcal{F}(\Delta E)}{T}= \lim_{\epsilon\to 0^{+}}\int_{-\infty}^{\infty}\, \DM u\, e^{-i\Delta E\,u} \, G^{+}_{\epsilon}\l(u\r). \nn \\
\end{eqnarray}
The pole structure of the Wightman function is given in Fig. 1 for rotational motion in different spacetimes.

\subsection{de Sitter spacetime}
\begin{figure*}
\includegraphics[width=5.50cm]{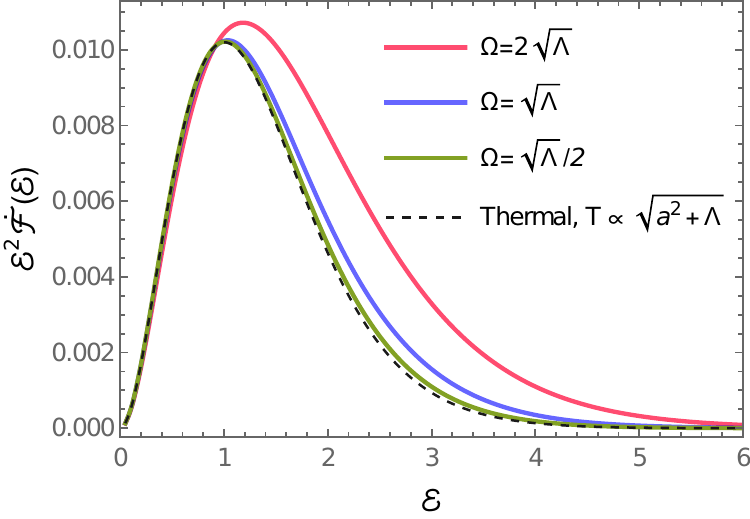}
\hskip 10pt
\includegraphics[width=5.50cm]{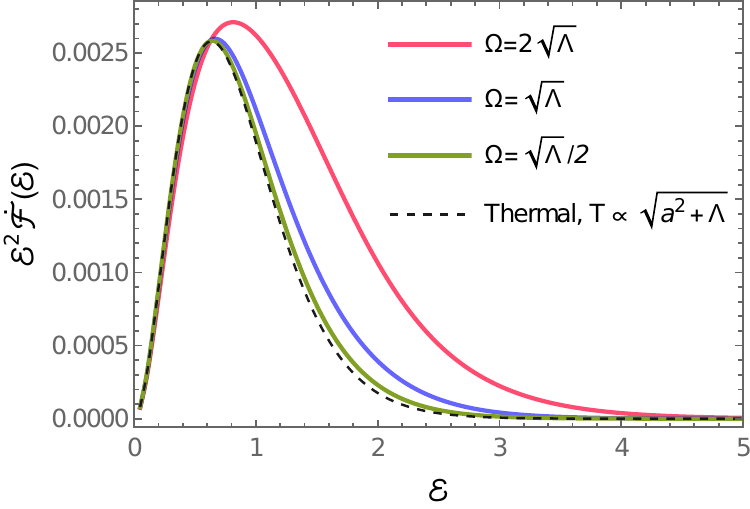}
\hskip 10pt
\includegraphics[width=5.50cm]{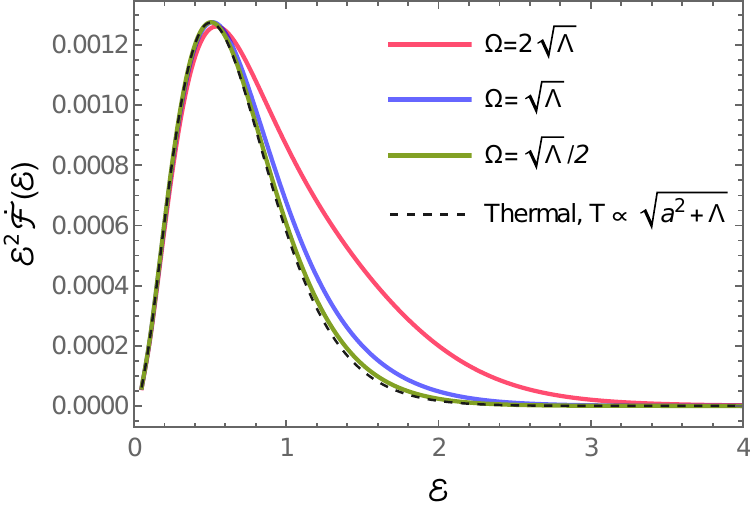}
\caption{The detector transition rate for a rotating detector in 3+1 de Sitter spacetime is plotted against the dimensionless energy gap, $\Delta E\,\sqrt{\Lambda}=\mathcal{E}$ for different values of acceleration and torsion. Left: $a=2\sqrt{\Lambda}$, Middle: $a=\sqrt{\Lambda}$, Right: $a=\sqrt{\Lambda}/2$. The dashed line corresponds to the thermal spectrum for a linearly accelerating detector in de Sitter.}
    \label{fig:deSitter-response}
\end{figure*}

In de Sitter spacetime we consider a conformally coupled scalar field for simplicity. The Wightman function for the conformally coupled scalar field\cite{Allen:1985ux,Garbrecht:2004du} is given by,
\begin{eqnarray}
    G^{+}\l(x,x^{\prime}\r)=-\frac{\Lambda}{4\pi^2 y_{\varepsilon}(x,x^{\prime})}
\end{eqnarray}
where, $y_\varepsilon(x,x^{\prime})$ is the de Sitter invariant distance function with proper $i\varepsilon$ prescription, $y(x,x^{\prime})\rightarrow y(\Delta\tau_{\text{geod}})=-4\sinh^2\l(\sqrt{\Lambda}\,\Delta\tau_{\text{geod}}/2\r)$. The transition rate is obtained using numerical methods and is shown in \fig{fig:deSitter-response} for different values of acceleration, torsion and curvature length scale. The transition rate approaches the thermal spectrum with temperature, $k_{B}T=\l(\hbar/2\pi\r) \sqrt{a^{2}+\Lambda}$ as $\Omega\rightarrow0$. 
%
%
\subsection{Anti-de Sitter spacetime}
\begin{figure*}
\includegraphics[width=5.50cm]{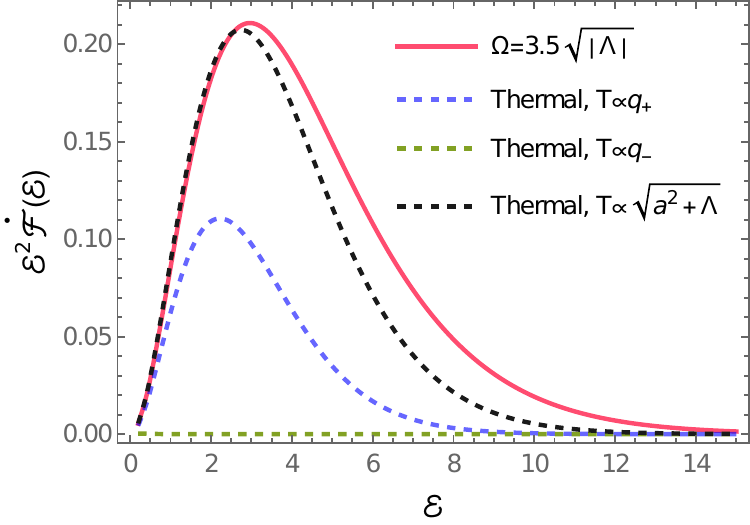}
\hskip 10pt
\includegraphics[width=5.60cm]{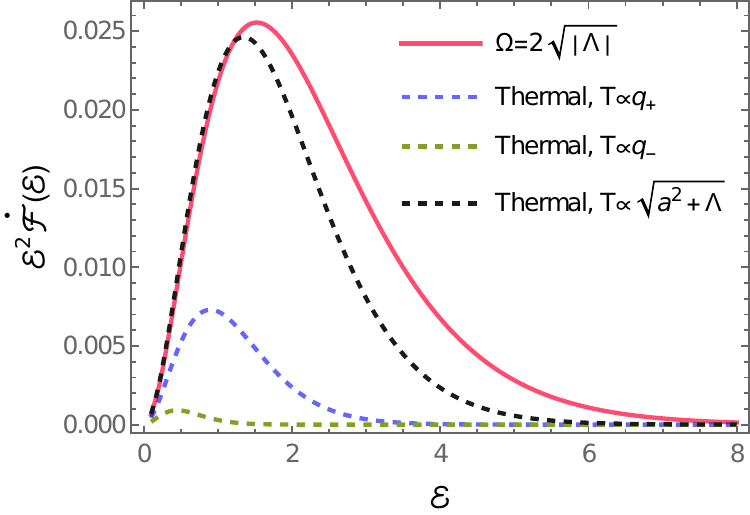}
\hskip 10pt
\includegraphics[width=5.55cm]{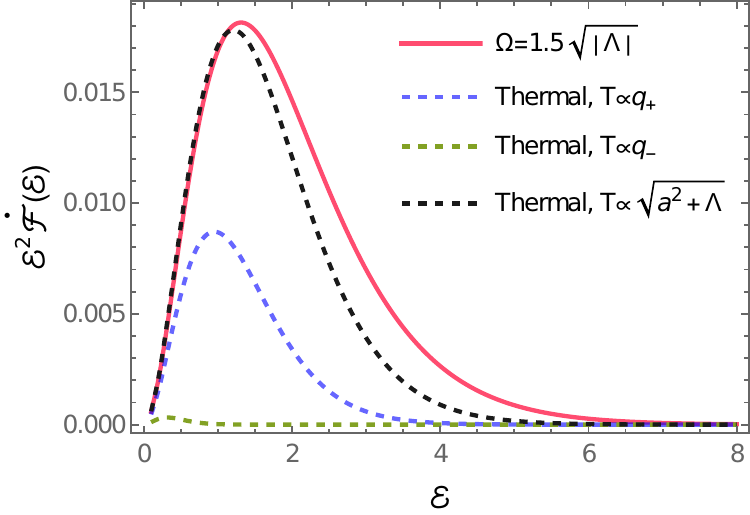}
\caption{
The detector transition rate for 3+1 Anti-de Sitter spacetime with transparent boundary condition plotted against the dimensionless energy gap, $\Delta E\,\sqrt{|\Lambda|}=\mathcal{E}$. Left: $a=1.5\sqrt{17}\sqrt{|\Lambda|}$, Middle: $a=\sqrt{10\,|\Lambda|}$, Right: $a=\sqrt{33\,|\Lambda|}/2$. All the parameter values in this plot satisfy the periodicity condition described in \ref{sec:periodicity}. The dashed lines corresponds to thermal spectrum with different temperature and it is clear that the detector transition rate do not correspond to thermal spectrum even though there is periodicity in geodesic distance for $\tau\to i\,t$.}
\label{fig:anti-deSitter-response}
\end{figure*}

For anti-de Sitter spacetime, we again consider a conformally coupled scalar field. The Wightman function is given by\cite{Avis:1977yn},
\begin{eqnarray}
    G^{+}_{\text{AdS}_4}\l(x,x^{\prime}\r)=-\frac{\Lambda}{4\pi^2}\l[\frac{1}{y(x,x^{\prime})}-\frac{\zeta}{y(x,x^{\prime})-2}\r]
\end{eqnarray}
where $\zeta\in \{0,-1,1\}$ corresponds to whether the boundary condition specified at infinity is transparent, Dirichlet or Neumann respectively. We will restrict our analysis to transparent boundary conditions for simplicity. In \fig{fig:anti-deSitter-response}, the response rate of an Unruh-DeWitt detector in stationary motion with acceleration and torsion with the parameter values chosen for periodicity in geodesic distance when $\tau\to it$ is illustrated. The transition rate do not correspond to
the thermal spectrum due to the presence of complex poles with real parts (see Fig. 1).
%
%

\section{Conclusions and Discussion}

We have derived some exact results for uniformly accelerated and rotating detectors in maximally symmetric spacetimes. The key result on which all others are based is the relation between geodesic and proper time intervals between two points on the detector trajectory. Since two-point functions, such as the Wightman function, depend on the former, whereas the detector response involves Fourier transform with respect to the latter, the analytic structure of this relation in the complex (proper time) plane determines the response. In particular, we uncover a periodicity in Euclidean time for certain values of rotation parameter $\Omega$ for $\Lambda<0$, while no such periodicity can be obtained for $\Lambda>0$ or $\Lambda=0$ (the well known case of uniformly rotating detector in Minkowski spacetime).
While periodicity of Wightman function in Euclidean time is one of the conditions for thermality, it is not the only one. There are certain analyticity conditions which are also required, as discussed in Ref. \cite{Fewster:2016ewy}; in particular, $|G^{+}(u)|$ should be bounded by a polynomial, $P(|\text{Re}\,u|)$ in the strip, $S=\{u\in\mathbb{C}\,|-\beta<\text{Im}\,u<0\}$, where $\beta$ is the periodicity. This condition is not satisfied for generic rotational motion due to the presence of additional complex poles with non-zero real parts.

Our results are important as they highlight the subtle role which curvature, howsoever small, can play in determining the \textit{nature} of response of accelerated, rotating detectors even in the point limit, where equivalence principle would generically imply only perturbative corrections.

In the spirit of the uniformly accelerated (rectilinear) motion discussed in \cite{hari:2021gns}, we also obtained perturbative expansions in curvature for accelerated, rotating detectors in arbitrary curved spacetime (ignoring the derivatives of curvature). However, unlike in \cite{hari:2021gns}, we have not been able to uncover a subset of terms that can be summed to an analytic expression. One obvious reason for this is the complexity of the series in presence of rotation. Besides, our exact results in maximally symmetric spacetimes offer no insights in presence of rotation which brings in an additional spacelike direction that, in turn, brings in more combination of curvature components than can be resolved by looking at dS/AdS, which has just one. We hope, however, to address this in a future work. Any re-summation involving acceleration, curvature and rotation is bound to be of tremendous significance not only in the context of quantum detectors, but also for classical processes in curved spacetimes that involve rotation.

\begin{acknowledgments}
H.K. would like to thank Indian Institute of Technology, Madras, Chennai, India and the Ministry of Human Resources and Development (MHRD), India for financial support. We would like to thank Prof. Jorma Louko for useful discussions and suggestions, and for drawing our attention to Ref. \cite{Fewster:2016ewy}.
\end{acknowledgments}

\appendix
\section{Minkowski spacetime} 
\label{stat-mink}

In Minkowski spacetime, the six stationary motions for different values of curvature constants are: i) uniform linear acceleration, ii) circular, iii) cusped, iv) catenary, and v) helical worldlines. 
Once the trajectory is expressed in terms of the tetrads at the origin, the relation between the geodesic distance and the proper time can be established using, $\Delta\tau^2_{\text{geod}}=\eta_{ab}x^a(\Delta\tau_{\text{traj}})x^b(\Delta\tau_{\text{traj}})$, where the arc length(proper time), $\tau$ is denoted by $\Delta\tau_{\text{traj}}$. Evaluating the derivatives in the Taylor expansion in terms of $u^i,n^i,b^i,d^i$ gives,
\begin{eqnarray}
\frac{\DM x^i}{\DM \tau}&=&u^i\nonumber\\
\frac{\DM^2 x^i}{\DM \tau^2}&=&a n^i \nonumber\\
\frac{\DM^3 x^i}{\DM \tau^3}&=&a \Omega b^i + a^2 \frac{\DM x^i}{\DM \tau}\nonumber\\
\frac{\DM^4 x^i}{\DM \tau^4}&=&a \Omega \lambda d^i + (a^2-\Omega^2) \frac{\DM^2 x^i}{\DM \tau^2}\nonumber\\
\frac{\DM^{\sf p} x^i}{\DM \tau^{\sf p}}&=&(a^2-\Omega^2-\lambda^2) \frac{\DM^{{\sf p}-2} x^i}{\DM \tau^{{\sf p}-2}}+a^2 \lambda^2 \frac{\DM^{{\sf p}-4} x^i}{\DM \tau^{{\sf p}-4}}\nonumber 
\end{eqnarray}
for ${\sf p} \ge 5$. Solving the above derivatives using the recurssion relations, we solved the odd and even derivative separately. The series expansion for the trajectory will sum to hyperbolic functions. The trajectory obtained using the solution of the recurssion relation after the summation is given as,
\begin{widetext}
\begin{eqnarray}
    x^i(\tau)&=&\frac{2}{q_{-}q_{+}q_{0}^2}\left[(a^2+\lambda^2+\Omega^2+q_{0}^2)q_{-}\sinh{\left(q_{+}\tau\right)}-(a^2+\lambda^2+\Omega^2-q_{0}^2)q_{+}\sinh{\left(q_{-}\tau\right)}\right]\,u^i(0)
    \nonumber\\
    &&+\frac{a\,\Omega}{q_{-}q_{+}q_{0}^2}\left[q_{-}\sinh{\left(q_{+}\tau\right)}-q_{+}\sinh{\left(q_{-}\tau\right)}\right]\, b^i(0) 
    \nonumber\\
    &&+\frac{\Omega}{a\, \lambda\, q_{0}^2}\left[q_{+}^2\left(\cosh{\left(q_{-}\tau\right)}-1\right)-q_{-}^2\left(\cosh{\left(q_{+}\tau\right)}-1\right)\right] \, d^i(0) 
    \nonumber\\
    &&+\frac{1}{2a q_{0}^2}\left[(a^2+\lambda^2+\Omega^2+q_{0}^2)\left\{\cosh{\left(q_{+}\tau\right)-1}\right\}-(a^2+\lambda^2+\Omega^2-q_{0}^2)\left\{\cosh{\left(q_{-}\tau\right)-1}\right\}\right]\, n^i(0)
\label{mink-traj}
\end{eqnarray}    
\end{widetext}
Here, $q_{0}:=\left[(a^2-\Omega^2-\lambda^2)^2+4a^2\lambda^2\right]^{1/4}$, $q_{\pm}:=(1/\sqrt{2})\sqrt{(a^2-\Omega^2-\lambda^2)\pm q_{0}^2}$. To find the relation between proper time of the trajectory and the geodesic distance, use $\Delta\tau^2_{\text{geod}}=\eta_{ab}x^a(\Delta\tau_{\text{traj}})x^b(\Delta\tau_{\text{traj}})$, noting that $u^i$, $n^i$, $b^i$, and $d^i$ are unit vectors and orthogonal to each other. Then the geodesic distance is,
\begin{widetext}
\begin{eqnarray}
\Delta\tau^2_{\text{geod}}= \frac{4}{q_{0}^2}\left[\left(1+\frac{(\Omega^2+\lambda^2)}{q_{+}^2}\right)\sinh^2{}\left(\frac{q_{+} \Delta\tau_{\text{traj}}}{2}\right)-\left(1+\frac{(\Omega^2+\lambda^2)}{q_{-}^2}\right)\sinh^2{}\left(\frac{q_{-} \Delta\tau_{\text{traj}}}{2}\right)\right]
\label{mink-sigma-sqr}
\end{eqnarray}
\end{widetext}

The above equation is valid for any general stationary motion in Minkowski spacetime. Special cases, such as hyperbolic motion \cite{Rindler:1960zz}, is easily obtained by putting torsion and hypertorsion to zero. Other trajectories such as circular motion can be obtained in a similar manner. The different stationary trajectories are as given below:

\subsection{Uniform linear acceleration}
For uniform linear acceleration or hyperbolic motion, the torsion and hypertorsion will be zero, hence the motion will be in the $u^i-n^i$ plane. The constants $q_{0},\,q_{-},\,q_{+}$ will reduce to $a,\,0,\,\sqrt{2}a$ respectively. The trajectory will be given by,
\begin{eqnarray}
x^i(\tau)&=&\frac{\sinh{\left(a\tau\right)}}{a}u^i(0)+\frac{\left[\cosh{\left(a\tau\right)-1}\right]}{a}n^i(0)
\label{mink-lin-acc-traj}
\end{eqnarray}
The equation for the geodesic distance from the above trajectory simplies to the well know result,
\begin{eqnarray}
\Delta\tau^2_{\text{geod}}= \frac{4}{a^2}\sinh^2\left(\frac{a}{2}\Delta\tau_{\text{traj}}\right)
\label{mink-lin-acc-sigma-sqr}
\end{eqnarray}

\subsection{Motion with acceleration and torsion} \label{mink-aw}
When hypertorsion vanishes and only acceleration and torsion are present, the trajectory will differ according to the values of $a$ and $\Omega$. The general motion with only $a$ and $\Omega$ motion can be expressed using the unit vectors, $u^i(0)$, $n^i(0)$, and $b^i(0)$. The constants $q_{0}$, $q_{-}$, and $q_{+}$ to $\sqrt{a^2-\Omega^2}$, $0$, and $\sqrt{a^2-\Omega^2}$ respectively. 
Applying this to \eq{mink-sigma-sqr} gives the relation between geodesic distance and proper time. In the case of stationary motion with no hypertorsion, the trajectories are classified into 3 types depending on the magnitudes of acceleration and torsion.

a)  \emph{Circular motion}: When $|\Omega|>|a|$ the motion is bounded(circular). The spatial projection of this motion is circular motion. 

b) \emph{Cusped motion}: For equal magnitudes of acceleration and torsion, the spatial projection of the motion is a cusp.

c) \emph{Catenary motion}: The spatial projection is catenary when the magnitude of the acceleration is greater than the torsion. The motion is unbounded and the equations for the trajectory and geodesic distance can be obtained by putting $\lambda=0$ in \eq{mink-traj} and \eq{mink-sigma-sqr} respectively. 

\subsection{Motion with acceleration, torsion and hypertorsion}
\label{mink-awl}
The equation for the trajectory is already given in \eq{mink-traj} and the geodesic distance is given in \eq{mink-sigma-sqr}. The equation for the geodesic distance of this general stationary motion is similar to stationary motion in maximally symmetric spacetimes with acceleration and torsion. The spatial projection of this motion is a helix.

\section{More on arbitrary curved spacetimes}\label{sec:exp-general}
More generally, the geodesic interval for a uniformly accelerated observer in arbitrarily curved spacetime can be represented as
\begin{eqnarray}
    \Delta \tau_{\text{geod}}^2 = F(a, R_{0n0n}) + \widetilde{F}(\nabla R_{abcd}) + \mathscr{R}_A
\end{eqnarray}
The first term is the partially resummed analytic part. As long as $R_{0n0n}$ is nearly constant, this part should be exact. Then any corrections for non-maximally symmetric spacetime will come from derivatives of the Riemann tensor and the term $\mathscr{R}_A$. 

For the part, $\mathscr{R}_A$ the terms can be written in the form of a series given by the recursion relation,
\begin{eqnarray}
    \mathscr{R}_A=\sum_{i=1}^{\infty}(\tau_{\rm acc})^{2i+6}\left[\sum_{j=0}^{i-1}a^{2(i-j)}\left\{\sum_{k=1}^{j+1}c_k R_A^{(k+1)}\mathscr{E}^{(1+j-k)}\right\}\right] \nn \\
\end{eqnarray}
Here, $R_A$ corresponds to the Riemann tensor with transverse component and $\mathscr{E}$ corresponds to the tidal tensor and $c_k$ is the constant coefficient. For comparing the contribution from each term using the appropriate length scales, the same can be represented as, 
\begin{eqnarray}
    \mathscr{R}_A=\frac{1}{a^2}\sum_{i=1}^{\infty}(\varepsilon_{\tau})^{2i+6}\left[\sum_{j=0}^{i-1}\left\{\sum_{k=1}^{j+1}c_k \left(\frac{\varepsilon_A}{\varepsilon_{0n}}\right)^{(k+1)}{(\varepsilon_{0n})}^{(2+j)}\right\}\right] \nn \\
\end{eqnarray}
where, $\varepsilon_{\tau}:=a\tau_{\rm acc}$, $\varepsilon_{A}:=R_A/a^2$ and $\varepsilon_{0n}:=\mathscr{E}/a^2$. As long as the curvature length scale along the direction of $\hat n$ is small, then ${\varepsilon_A}/{\varepsilon_{0n}}$ will be small and the contribution from $\mathscr{R}_A$ will be subdominant.

The contribution from the terms $\widetilde{F}(\nabla R_{abcd})$, can be small as long as the length scale corresponding to such terms is larger than the length scale of acceleration and $\mathscr{E}_n$ term. For clarity, consider the eight-order terms in the series (derivative terms starts at order $\mathcal{O}(\tau^7)$), will have the structure: $a^2\nabla_{0}\nabla_{0}\mathscr{E}, a^3\nabla_{n}\mathscr{E}, a^4\mathscr{E}, a^{2}\mathscr{E}^2$ and $a^6$. Demanding the gradients of Riemann to be small at this order, leads to the conditions: $\nabla_{0}\nabla_{0}\mathscr{E}_n/(a^2\mathscr{E}_n)\ll 1$, $\nabla_{0}\nabla_{0}\mathscr{E}_n/\mathscr{E}^2_n\ll 1$, $\nabla_{n}\mathscr{E}_n/(a\mathscr{E}_n)\ll 1$ and $a\nabla_{n}\mathscr{E}_n/\mathscr{E}_n^2\ll 1$ at this order. As long as the length scale of acceleration and $\mathscr{E}_n$ are much smaller than the gradients of Riemann, all corrections due to derivative are also subdominant.

\bibliographystyle{apsrev}



\end{document}